\documentstyle[aps,prb]{revtex}

\begin{document}
\draft
\title{ Exactly solvable Kondo problem in the open $t-J$ chains}

\author{Jianhui Dai$^{a,b}${~~~} Yupeng Wang$^{c,d}$}
\address{
$^a$Abdus Salam International Center for Theoretical Physics, Trieste 34100, Italy\\
$^b$Zhejiang Institute of Modern Physics, Zhejiang University, Hangzhou 310027, China\\
$^c$Institut f\"{u}r Physik, Universit\"{a}t Augsburg, D-86135, Augsburg, Germany\\
$^d$Cryogenic Laboratory, Chinese Academy of Sciences, Beijing 100080,  China}

\maketitle
\begin{abstract}
We study the problem  of a boundary magnetic  impurity coupled with the solvable $t-J$ chains. Our model provides a good
start point to understand the Kondo problem in a Luttinger liquid as well as in a strongly correlated host. The Kondo
coupling constant $J$ can take arbitrary values, which allows us to study the ferromagnetic and antiferromagnetic Kondo
problems simultaneously. It is shown 
that both the Kondo coupling and the scalar potential effectively reconcile the spin dynamics at low temperatures. 
Generally, the impurity spin
is split into two ghost spins via coupling effect. The residual entropy, which can be exactly derived for the present model,
is strongly interaction-dependent. This hints the local spin configuration is very complicated and very different from
that in the conventional Kondo problem. An unscreening phenomenon in the antiferromagnetic regime is found, which reveals
the impurity potential plays an important role for the Kondo problem in a strongly correlated host. The low temperature
specific heat is calculated in the framework of local Landau-Luttinger liquid theory.
\end{abstract}

\section{introduction}
With the development of the nanofabrication techniques for quantum wires and the prediction of edge states in the quantum 
Hall effect, the interest in one-dimensional (1D) quantum systems has been renewed in recent years\cite{1,2}. In fact, much
of the interest in 1D quantum systems is due to Anderson's observation\cite{3} that the normal state properties of the
quasi 2D high-$T_c$ superconductors are strikingly different from all known metals and can not be reconciled with Landau's
Fermi liquid theory but are more similar to properties of 1D metals. In another hand, the impurity problem has been a current 
interest in the field of condensed matter physics. A well known example is the Kondo problem, which stimulated a strong
challenge to traditional perturbation theory and provided a possible ``laboratory" to search for the non-Fermi liquid behavior.
The local perturbation problem to a 1D Fermi system has been the subject of an intensive theoretical investigation in the
recent years, for both its interesting anomalies with respect to that of a higher-dimensional system, and its relevance to
a variety physical situations such as the transport behavior of the quantum wires\cite{2,4} and the tunneling through a
constriction in the quantum Hall regime\cite{5}. Kane and Fisher\cite{6} have argued that a single impurity in a 1D repulsive
interacting system in fact corresponds to a chain disconnected at the impurity site at low energy scales. Their observation
was also justified in the framework of the renormalization group analysis of boundary conformal field theory\cite{7,8},
which show that the open boundary condition is the stable fixed point for repulsive interaction. An interesting problem is the 
1D electron system couples with a magnetic impurity or the Kondo problem in a 1D metals. This problem was first considered by 
Lee and Toner\cite{9}, who found the crossover of the Kondo temperature from power law dependence on the Kondo coupling
constant to an exponential one, when the electron correlation goes from strong limit to weak limit. Subsequently, a poor
man's scaling was performed by Furusaki and Nagaosa\cite{10}, who addressed a conjecture that ferromagnetic Kondo screening
may occur in 1D due to the special topology. The boundary conformal field theory\cite{11}
gave out a classification of critical behavior for the 1D Kondo problem (without impurity potential). It turns there are 
only two possibilities, a standard low-temperature thermodynamics or a non-Fermi liquid observed by Furusaki and Nagaosa.
The non-Fermi liquid behavior is induced by the tunneling effect of conduction electrons through the impurity which depends
only on the bulk properties but not on the detail of the impurity\cite{12}.
\par
Despite such important progress, the problem of few impurities (potential, magnetic, especially both) embedded in a strongly
correlated 1D system is still not very well understood. We remark the study on the integrable models generally gives some
useful information to understand a variety of physical situations in 1D, if not fully. There are a few integrable models
related to the impurity problem in 1D quantum systems: an impurity spin embedded in a spin-$1/2$ Heisenberg chain solved 
many years ago by Andrei and Johannesson\cite{13}; an integrable impurity in the supersymmetric $t-J$ model\cite{14} solved
by Bed\"{u}rftig et al.. The impurities in these models are exactly transparent due to the unphysical terms in the hamiltonians
and contradict to the predictions of Kane and Fisher\cite{6}. This shortcoming was overcome in some other integrable models\cite{15,16,17}
by introducing the open boundary condition at the impurity site. Unlike in 3D, the scalar potential\cite{15} and the bond
deformation\cite{17} have significant effects to the low-temperature thermodynamics of the impurity. The anomaly residual
entropy induced by the boundary effect was predicted in these models, which strongly suggests that the charge-spin cooperation
in 1D correlated electron systems may play an important role.
\par
In this paper, we consider a boundary magnetic impurity coupled with the integrable $t-J$ chains. This is the detailed and extended
description of our earlier letter\cite{15}. The Hamiltonian we shall study reads
\begin{eqnarray}
H=H_0+H_i, \nonumber\\
H_0=-t\sum_{j=1}^{L-1}\sum_\sigma(c_{j,\sigma}^\dagger c_{j+1,\sigma}+h.c.)+\sum_{j=1}^{L-1}(J{\bf S}_j\cdot{\bf S}_{j+1}
+Vn_jn_{j+1}),\\
H_i=2J_i{\bf S}_L\cdot{\bf S}+V_in_L,\nonumber
\end{eqnarray}
where $L$ is the length or site number of the chain; $c_{j,\sigma}^\dagger$ ($c_{j,\sigma}$) are the creation (annihilation)
operators of the conduction electrons and ${\bf S}_j=1/2\sum_{\sigma,\sigma'}c_{j,\sigma}^\dagger{\bf\tau}_{\sigma,\sigma'}
c_{j,\sigma'}$ are the spin operators of the conduction electrons; $n_{j,\sigma}=\sum_{\sigma}c_{j,\sigma}^\dagger c_{j,\sigma}$
is the number operator; ${\bf S}$ is the impurity spin with strength $S$; $t,J,V, J_i$ and $V_i$ are all constants. Notice that
the single occupation condition $n_j\leq 1$ is assumed in (1). Without losing generality, we shall put $t=1$ in the following
text.
\par
The structure of the present paper is the following: In the subsequent section, we derive the integrable conditions for a
variety of choices of the parameters $J$ and $V$. The reflection equation \cite{18} constrains the integrability of the
present model. That means the parameters $J_i$ and $V_i$ can not be independent but are parameterized by an unique c-number
$c$ to ensure the integrability of the model. In addition, the Bethe ansatz equations for the integrable cases will be
given. In sect. III, we study the ground state properties and low-lying excitations. The formation of boundary bound states
will be discussed in detail. Sect.IV is attributed to the derivation of the thermodynamic equations. The residual entropy
induced by the impurity will be discussed. A different method, i.e., the local Landau-Luttinger liquid
description\cite{16,17} is used to study the low-temperature thermodynamics of the impurity. Such a method can also be applied to non-integrable
models and is thus general to the impurity problem in the 1D quantum systems. Concluding remarks will 
be given in Sect.V.
In appendix A, the general solution for the reflection matrix is given. The appendix B is attributed to the eigenvalue problem
of the nested Bethe ansatz.
\section{Integrable conditions}
It is well known that $H_0$ is exactly solvable for $J=\pm2$, $V=\mp1/2,\pm3/2$\cite{19,20}. By including the impurity, any
electron impinging on the impurity will be completely reflected and suffer a reflection matrix $R_j$. The waves are
therefore reflected at either end as
\begin{eqnarray}
e^{ik_jx}\to -e^{-ik_jx}, {~~~}x\sim1\nonumber\\
e^{ik_jx}\to R_j(k_j)e^{-ik_jx-2ik_jL}, {~~~}x\sim L.
\end{eqnarray}
Let us consider the two particle case. There are two ways from an initial state $(k_1,k_2,|)$ to a final state $(-k_1,-k_2,
|)$:
$$
I. (k_1,k_2,|)\to(k_2,k_1,|)\to(k_2,-k_1,|)\to(-k_1,k_2,|)\to(-k_1,-k_2,|),
$$
$$
II. (k_1,k_2,|)\to(k_1,-k_2,|)\to(-k_2,k_1,|)\to(-k_2,-k_1,|)\to(-k_1,-k_2,|),
$$
where the symbol $|$ denotes the open boundary. Since the physical process is unique, the following equation must hold
\begin{eqnarray}
S_{12}(k_1,k_2)R_1(k_1)S_{12}(k_1,-k_2)R_2(k_2)=R_2(k_2)S_{12}(k_1,-k_2)R_1(k_1)S_{12}(k_1,k_2).
\end{eqnarray}
Above $S_{12}$ is the scattering matrix between the two electrons. This is just the reflection equation
\cite{18}. For the multi-particle cases,
as long as the scattering matrix is factorizable or the two-body scattering matrix satisfies 
the Yang-Baxter relation\cite{21}
\begin{eqnarray}
S_{12}(k_1,k_2)S_{13}(k_1,k_3)S_{23}(k_2,k_3)=S_{23}(k_2,k_3)S_{13}(k_1,k_3)S_{12}(k_1,k_2),
\end{eqnarray}
(3) is the only constraint to the integrability of an open boundary system\cite{18}. Normally, a c-number reflection matrix indicates either a
boundary field in the spin chain models or a scalar potential in a fermion system. However, in the present model, the boundary impurities have internal 
degrees of freedom and the spin-exchange processes must be included when an electron is reflected by the boundary. That means the
reflection matrix $R_j$ must be an operator matrix rather than a c-number one. Below we derive the integrable conditions
for the different choices of $J$ and $V$.
\par
(i)$J=2, V=3/2$ case. Since the reflection process only consists of one-electron effect, it is convenient to consider
the single-particle eigenstate. The Schr\"{o}dinger equation for this case reads
\begin{eqnarray}
-\Psi(x+1)-\Psi(x-1)=E\Psi(x), {~~~~}1<x<L\\
-\Psi(2)=E\Psi(1), {~~~~}x=1\\
-\Psi(L-1)+(V_i+J_i\sigma\cdot{\bf S})\Psi(L)=E\Psi(L), {~~~~}x=L.
\end{eqnarray}
We make the following ansatz for the wave function $\Psi(x)$
\begin{eqnarray}
\Psi(x)=A_+e^{ikx}+A_-e^{-ikx},
\end{eqnarray}
then the eigenvalue $E$ takes the value $-2\cos k$. From (6) we obtain
\begin{eqnarray}
\frac{A_+}{A_-}=-1.
\end{eqnarray}
Similarly, from (7) we obtain
\begin{eqnarray}
R(k)=\frac{A_-}{A_+}e^{-2ikL}=-\frac{e^{ik}+(V_i+J_i\sigma\cdot{\bf S})}{e^{-ik}+(V_i+J_i\sigma\cdot{\bf S})}.
\end{eqnarray}
The self-consistent condition for $A_\pm$ constrains the value of $k$ by the following eigenvalue problem
\begin{eqnarray}
-R(k)A_-=e^{-2ikL}A_-.
\end{eqnarray}
For arbitrary $N-$particle case, $R_j(k_j)$ must satisfy the reflection equation (3). It is known that the two-body scattering matrix takes
the form
\begin{eqnarray}
S_{jl}(q_j-q_l)=-\frac{q_j-q_l+iP_{jl}}{q_j-q_l-i},
\end{eqnarray}
where $q_j=1/2tan(k_j/2)$, $P_{jl}$ is the spin exchange operator\cite{19}.
\par
As shown in appendix A, the reflection matrix $R_j(k_j)$ must take the following form
\begin{eqnarray}
R_j(k_j)=e^{i\varphi(q_j)}\frac{q_j-ic+i(\sigma_j\cdot{\bf S}+1)}{q_j-ic+i(S+1)}\frac{q_j+ic+i\sigma_j
\cdot{\bf S}}{q_j+ic+iS},
\end{eqnarray}
where $c$ is an arbitrary constant. By virtue of the relation
\begin{eqnarray}
(\sigma_j\cdot{\bf S})^2=S(S+1)-\sigma_j\cdot{\bf S},
\end{eqnarray}
we obtain  $\sigma_j\cdot{\bf S}$ has two eigenvalues. Those are $S$ and $-(S+1)$. For $\sigma_j\cdot{\bf S}=S$, comparing
(10) and (13) we have
\begin{eqnarray}
e^{i\varphi(k_j)}=-\frac{e^{ik_j}+(V_i+SJ_i)}{e^{-ik}+(V_i+SJ_i)}.
\end{eqnarray}
For $\sigma_j\cdot{\bf S}=-(S+1)$, we have
\begin{eqnarray}
\frac{q_j-ic-iS}{q_j+ic+iS}\frac{q_j+ic-i(S+1)}{q_j-ic+i(S+1)}=\frac{e^{ik_j}+[V_i-(S+1)J_i]}{e^{-ik}+[V_i-(S+1)J_i]}
\frac{e^{-ik_j}+(V_i+SJ_i)}{e^{ik}+(V_i+SJ_i)}.
\end{eqnarray}
The solution of the above equation for $V_i$ and $J_i$ leads to the integrable condition
\begin{eqnarray}
J_i=\frac1{(S+\frac 12)^2-c^2},{~~~~~}V_i=\frac{S^2+S+1-(c-\frac12)^2}{(S+\frac12)^2-c^2},
\end{eqnarray}
 or
\begin{eqnarray}
J_i=\frac1{(S+\frac 12)^2-(1-c)^2},{~~~~~}V_i=\frac{S^2+S+1-(c-\frac12)^2}{(S+\frac12)^2-(1-c)^2}.
\end{eqnarray}
Obviously, to ensure the integrability of the present model, $J_i$ and $V_i$ must be related to each other 
and parameterized by a unique constant $c$ as shown in (17) and (18). Notice that (18) is dual to the solution of (17) under the shift $c\to 1-c$. Without losing generality, we use (17) in the following text. The reflection matrix (10)
in the integrable case can be rewritten as
\begin{eqnarray}
R_j(q_j)=\frac{q_j-\frac i2}{q_j+\frac i2}\frac{q_j-ic+i(\sigma\cdot{\bf S}+1)}{q_j+ic+i(S+1)}\frac{q_j+ic+i\sigma\cdot
{\bf S}}{q_j+ic+iS}.
\end{eqnarray}
For an $N-$particle system, suppose the wave function initially has an amplitude $\zeta_0$. When the $j-th$ particle moves 
across another particle, it gets an $S-$matrix $S_{jl}(q_j-q_l)$. At the right boundary, it is completely reflected back and suffered a factor
$\exp(2ik_jL)R_j(q_j)$. Then it begins to move toward the left boundary. When it reaches the left boundary, it will be kicked back and suffered
a factor $-1$. Finally it arrives at the initial site and finishes a periodic motion. Therefore we have the following
equation
\begin{eqnarray}
-S_{jj-1}^-\cdots S_{j1}^-S_{j1}^+\cdots S_{jj-1}^+S_{jj+1}^+\cdots S_{jN}^+R_je^{2ik_jL}S_{jN}^-\cdots S_{jj+1}^-\zeta_0=\zeta_0,
\end{eqnarray}
or more neatly
\begin{eqnarray}
S_{jj-1}^-\cdots S_{j1}^-S_{j1}^+\cdots S_{jj-1}^+S_{jj+1}^+\cdots S_{jN}^+R_jS_{jN}^-\cdots S_{jj+1}^-\zeta_0=-
(\frac{q_j+\frac i2}{q_j-\frac i2})^{2L}\zeta_0,
\end{eqnarray}
where $S_{jl}^\pm=S_{jl}(q_j\pm q_l)$. Eq.(21) is just the reflection version of Yang's eigenvalue problem\cite{21}. Its
solution gives out
 the Bethe ansatz equation. To keep the continuum of the text, we attribute the solution of (21) to appendix B.
Below we write down the result
\begin{eqnarray}
(\frac{q_j-\frac i2}{q_j+\frac i2})^{2L+1}=-\frac{q_j-i(S+1-c)}{q_j+i(S+1-c)}\prod_{r=\pm}\prod_{l\neq j}^N
\frac{q_j-rq_l-i}{q_j-rq_l+i}\prod_{\alpha=1}^M\frac{q_j-r\lambda_\alpha+\frac i2}{q_j-r\lambda_\alpha-\frac i2},
\end{eqnarray}
\begin{eqnarray}
\frac{\lambda_\alpha+i(S+\frac12-c)}{\lambda_\alpha-i(S+\frac12-c)}\frac{\lambda_\alpha+i(S-\frac12+c)}{\lambda_\alpha-i
(S-\frac12+c)}\prod_{r=\pm}\prod_{j=1}^N\frac{\lambda_\alpha-rq_j+\frac i2}{\lambda_\alpha-rq_j-\frac i2}
=\prod_{r=\pm}\prod_{\alpha\neq \beta}^M\frac{\lambda_\alpha-r\lambda_\beta+i}{\lambda_\alpha-r\lambda_\beta-i},
\end{eqnarray}
where $M\leq N/2$ is the number of spin-down electrons. The eigenenergy is given by
\begin{eqnarray}
E=2N-\sum_{j=1}^N\frac 4{4q^2_j+1}.
\end{eqnarray}
The derivation of the integrable conditions and the Bethe ansatz equations for other choices of $J$ and $V$ is similar. 
Here we omit the detail and only write down the result.
\par
(ii)$ J=2, V=-1/2$. In this case, the scattering matrix takes the form
\begin{eqnarray}
S_{jl}(q_j-q_l)=\frac{q_j-q_l-iP_{jl}}{q_j-q_l-i},{~~~~}q=\frac12\tan^{-1}\frac {k_j}2.
\end{eqnarray}
The integrable condition is
\begin{eqnarray}
J_i=\frac1{(S+\frac 12)^2-c^2},{~~~~~}V_i=\frac{S^2+S+1-(c+\frac12)^2}{(S+\frac12)^2-c^2},
\end{eqnarray}
and the reflection matrix is
\begin{eqnarray}
R_j(q_j)=-\frac{q_j+\frac i2}{q_j-\frac i2}\frac{q_j-ic-i(\sigma\cdot{\bf S}+1)}{q_j+ic+i(S+1)}\frac{q_j+ic-i\sigma\cdot
{\bf S}}{q_j+ic-iS}.
\end{eqnarray}
The Bethe ansatz equations read as
\begin{eqnarray}
(\frac{q_j+\frac i2}{q_j-\frac i2})^{2L+1}=\frac{q_j-i(S+1+c)}{q_j+i(S+1+c)}\prod_{r=\pm}
\prod_{\alpha=1}^M\frac{q_j-r\lambda_\alpha+\frac i2}{q_j-r\lambda_\alpha-\frac i2},
\end{eqnarray}
\begin{eqnarray}
\frac{\lambda_\alpha-i(S+\frac12-c)}{\lambda_\alpha+i(S+\frac12-c)}\frac{\lambda_\alpha-i(S-\frac12+c)}{\lambda_\alpha+i
(S-\frac12+c)}\prod_{r=\pm}\prod_{j=1}^N\frac{\lambda_\alpha-rq_j-\frac i2}{\lambda_\alpha-rq_j+\frac i2}
=\prod_{r=\pm}\prod_{\alpha\neq \beta}^M\frac{\lambda_\alpha-r\lambda_\beta-i}{\lambda_\alpha-r\lambda_\beta+i}.
\end{eqnarray}
\par
(iii)$J=-2,V=1/2$ case. In this case, the two-body scattering matrix takes the form
\begin{eqnarray}
S_{jl}(q_j-q_l)=\frac{q_j-q_l+iP_{jl}}{q_j-q_l+i}, {~~~~}q_j=\frac12\tan\frac{k_j}2.
\end{eqnarray}
The integrable condition is the same to that of case (i) and the Bethe ansatz equations are
\begin{eqnarray}
(\frac{q_j+\frac i2}{q_j-\frac i2})^{2L+1}=-\frac{q_j+i(S+1-c)}{q_j-i(S+1-c)}\prod_{r=\pm}
\prod_{\alpha=1}^M\frac{q_j-r\lambda_\alpha-\frac i2}{q_j-r\lambda_\alpha+\frac i2},
\end{eqnarray}
\begin{eqnarray}
\frac{\lambda_\alpha-i(S+\frac12-c)}{\lambda_\alpha+i(S+\frac12-c)}\frac{\lambda_\alpha-i(S-\frac12+c)}{\lambda_\alpha+i
(S-\frac12+c)}\prod_{r=\pm}\prod_{j=1}^N\frac{\lambda_\alpha-rq_j-\frac i2}{\lambda_\alpha-rq_j+\frac i2}
=\prod_{r=\pm}\prod_{\alpha\neq \beta}^M\frac{\lambda_\alpha-r\lambda_\beta-i}{\lambda_\alpha-r\lambda_\beta+i}.
\end{eqnarray}
(iv)$J=-2, V=-3/2$ case. In this case, the two-body scattering matrix is
\begin{eqnarray}
S_{jl}(q_j-q_l)=-\frac{q_j-q_l-iP_{jl}}{q_j-q_l+i},{~~~~}q_j=\frac 12\tan^{-1}\frac{k_j}2.
\end{eqnarray}
The integrable condition is the same to that of case (ii) and the Bethe ansatz equations read as
\begin{eqnarray}
(\frac{q_j-\frac i2}{q_j+\frac i2})^{2L+1}=\frac{q_j-i(S+1+c)}{q_j+i(S+1+c)}\prod_{r=\pm}\prod_{l\neq j}^N
\frac{q_j-rq_l-i}{q_j-rq_l+i}\prod_{\alpha=1}^M\frac{q_j-r\lambda_\alpha+\frac i2}{q_j-r\lambda_\alpha-\frac i2},
\end{eqnarray}
\begin{eqnarray}
\frac{\lambda_\alpha+i(S+\frac12-c)}{\lambda_\alpha-i(S+\frac12-c)}\frac{\lambda_\alpha+i(S-\frac12+c)}{\lambda_\alpha-i
(S-\frac12+c)}\prod_{r=\pm}\prod_{j=1}^N\frac{\lambda_\alpha-rq_j+\frac i2}{\lambda_\alpha-rq_j-\frac i2}
=\prod_{r=\pm}\prod_{\alpha\neq \beta}^M\frac{\lambda_\alpha-r\lambda_\beta+i}{\lambda_\alpha-r\lambda_\beta-i},
\end{eqnarray}
In the following text, we concentrate ourselves on case (i). Other cases can be treated with the same procedure without
any difficulty.
\section{Ground state properties and low-lying excitations}
\subsection{Boundary bound state and zero temperature magnetization}
Unlike in the homogeneous systems, some boundary modes may exist in the present model due to the presence of the impurity.
From the Bethe ansatz equation (22) we can see an imaginary mode $q=i(S+1-c)$ is a possible solution as long as $c<S+1$.
However, such a mode is not always stable in the ground state. As we can read off from (24), the real modes form a band
$-2\leq \epsilon(q)\leq 2$. The energy of the imaginary mode falls either below the band or above it. Therefore, the
imaginary $q$ mode defines a true bound state around the impurity. Below we discuss the formation of the boundary bound 
state and zero temperature magnetization for different $c$ values.
\par
(i)$c\geq S+1$. In this case, the system falls into the regime of ferromagnetic Kondo coupling. There is no boundary bound
state of charges at low energy scales. For convenience, we put $N$ to be odd in this case. In the ground state, all the
modes $\{q_j\}$ and $\{\lambda_\alpha\}$ take real values. Define the quantities
\begin{eqnarray}
Z_L^c(q)=\frac1{2 L}\{\sum_{r=\pm}\sum_{\alpha=1}^M\theta_1(q-r\lambda_\alpha)-\sum_{r=\pm}\sum_{j=1}^N\theta_2(q-rq_j)+
\phi_c^e(q)+\phi_c^i(q)\}+\theta_1(q),\\
Z_L^s(\lambda)=\frac1{2L}\{\sum_{r=\pm}\sum_{j=1}^N\theta_1(\lambda-rq_j)-\sum_{r=\pm}\sum_{\beta=1}^M
\theta_2(\lambda-r\lambda_\beta)+\phi_s^e(\lambda)+\phi_s^i(\lambda)\}\nonumber
\end{eqnarray}
where $\theta_n=2\arctan(2x/n)$ and
\begin{eqnarray}
\phi_c^e(q)=2\theta_1(q),{~~~~}\phi_c^i(q)=\theta_{2(c-S-1)}(q),\\
\phi_s^e(\lambda)=\theta_2(\lambda){~~~~}\phi_s^i(\lambda)=\theta_{(2c+2S-1)}(\lambda)-\theta_{(2c-2S-1)}(\lambda).
\end{eqnarray}
Notice the zero modes are
forbidden in an open boundary system\cite{23}.  Obviously, $Z_L^c(q_j)=\pi I_j/L$, $Z_L^s(\lambda_\alpha)
=\pi J_\alpha/L$ give the logarithmic version of the Bethe ansatz equations, where $I_j$ and $J_\alpha$ are integers. In the 
ground state, both $\{I_j\}$ and $\{J_\alpha\}$ are closely packed numbers from $1$ up to $ N$ and $ M$ respectively.
The cutoffs of $q$ and $\lambda$ are defined as $Z_L^c(Q)=\pi(N+1/2)/L$, $Z_L^s(\Lambda)=\pi(M+1/2)/L$. Define the density
functions as
\begin{eqnarray}
\rho_L^c(q)=\frac 1{2\pi}\frac{dZ_L^c(q)}{dq}-\frac 1{2L}\delta(q),{~~~~}\rho_L^s(\lambda)=
\frac1{2\pi}\frac{dZ_L^s(\lambda)}{d\lambda}-\frac1{2L}\delta(\lambda).
\end{eqnarray}
We have the relation
\begin{eqnarray}
\int_{-Q}^Q\rho_L^c(q)dq=\frac 1LN,{~~~~}\int_{-\Lambda}^\Lambda\rho_L^s(\lambda)d\lambda=\frac 1LM.
\end{eqnarray}
As demonstrated by many authors\cite{24}, $\Lambda\to\infty$ for the ground state in the thermodynamic limit $L\to \infty$, since
any hole in the real $\lambda-$axis induces an excited state\cite{25}. From(40) we deduce that $N=2M$, which means the residual
magnetization is 
\begin{eqnarray}
M_s=S+\frac12N-M=S.
\end{eqnarray}
Therefore, the impurity moment can not be screened any more in this case.
\par
(ii)$S+1/2<c<S+1$. In this case, the system falls still in the ferromagnetic Kondo regime. Due to the stronger Kondo coupling
and the weaker impurity potential, one boundary bound state occurs with $q=i(S+1-c)$. Since $|q|<1/2$, we can see the energy carried
by this mode
\begin{eqnarray}
\epsilon(q)=2-\frac4{1-4(S+1-c)^2}<-2
\end{eqnarray}
is below the band. Therefore this state is much more stable than any other states. Similarly, we can define the functions 
$Z_L^c(q)$ and $Z_L^s(\lambda)$ for the real modes. In this case, the summation for the real charge modes runs from $1$ to
$N-1$ and $\phi_c^i(q)$ and $\phi_s^i(\lambda)$ are given by
\begin{eqnarray}
\phi_c^i(q)=-\theta_{2(S+1-c)}-\theta_{2(c-S)}(q)-\theta_{2(S+2-c)}(q),\\
\phi_s^i(\lambda)=\theta_{(2c+2S-1)}(\lambda)+\theta_{(2S+3-2c)}(\lambda).
\end{eqnarray}
Notice the phase shifts $\phi_{c,s}^e$ induced by the open boundary are not changed.
By integrating $\rho_L^s(\lambda)$, we obtain again $N=2M$. It seems that the local spin is still unscreened. However, we
remark the localized electron and the impurity form a spin-$(S+1/2)$ composite. When $c\to S+1/2+0^+$, $J,V\to -\infty$, the
composite behaves as a perfect local spin with spin-$(S+1/2)$. It is the composite rather than the original impurity interacting
with the host effectively. In such a sense, we can say the local moment is partially screened. If we include another half chain
interacting with the impurity, the problem becomes a two-channel Kondo problem\cite{26,27} and we expect the realization of Furusaki-Nagaosa's
conjecture.
\par
(iii)$-(S-1/2)<c<S+1/2$. The system falls into the antiferromagnetic Kondo regime. The boundary bound state $q=i(S+1-c)$ is no longer
a stable state since it contains much higher energy. Like in the $c>S+1$ case, the ground state is described by closely
packed real $q-$modes and $\lambda-$modes. The residual magnetization is  $S-1/2$, which indicates 
the  Kondo screening, like the case in the traditional Kondo problem.\par
(iv)$-S<c<-(S-1/2)$.
Notice that both $J_i$ and $V_i$ are very large in this case. It seems $V_i$ dominates over $J_i$ and prevents the conduction
electrons near to the impurity. The residual magnetization is $S$, which means there is no Kondo screening.
Such a situation is very different from that of a Kondo problem in a 3D Fermi liquid, where
the impurity potential does not change the fixed point and only induces the renormalization of the 
Kondo coupling constant\cite{28}.
Such a phenomenon strongly suggests the charge-spin cooperation plays an important role in the 1D Kondo problem.
\par
(v)$-(S+1/2)<c<-S$. In this case, the system is still in the regime of antiferromagnetic Kondo coupling. From the Bethe ansatz
equations(22-23) we can see an imaginary spin mode $\lambda=i(S-1/2+c)$ may assist the formation of a boundary bound state with
$q=i(S+c)$. This mode carries much lower energy than that of a real mode and is therefore stable in the ground state. The functions
$Z_L^c(q)$ and $Z_L^s(\lambda)$ can be defined in a similar way. Notice here the summation for the real charge modes runs
from $1$ to $N-1$ and that for the spin modes runs from $1$ to $M-1$. $\phi_c^i(q)$ and $\phi_s^i(\lambda)$ are defined as
\begin{eqnarray}
\phi_c^i(q)=-\theta_{2(1+S-c)}(q)-\theta_{2(S+c+1)}(q)-\theta_{2|S+c|}(q),\\
\phi_s^i(\lambda)=\theta_{(2S-2c+1)}-\theta_{|2S+2c-3|}(\lambda).
\end{eqnarray}
In the thermodynamic limit, we have $N=2M-1$ in the ground state. Therefore, the residual magnetization takes the value of
$S-1/2$, which indicates a typical Kondo screening. In fact, both the positive $J_i$ and negative $V_i$ assist one conduction
electron to form tight-bonding pair with the local moment. The bounded  spin mode is a signal that the local spin-$(S-1/2)$
pair is formed.
\par
(vi)$c<-(S+1/2)$. In this case, both the ferromagnetic Kondo coupling ($J_i>0$) and the repulsive impurity potential forbid
the electrons to form a bound state with the impurity. $Z_L^c(q)$ and $Z_L^s(\lambda)$ can be defined in the same forms as
(36) but with
\begin{eqnarray}
\phi_c^i(q)=-\theta_{2|S+1-c|}(q),\\
\phi_s^i(\lambda)=\theta_{(2S-2c+1)}(\lambda)-\theta_{|2S+2c-1|}(\lambda).
\end{eqnarray}
The residual magnetization is still $S$ and no Kondo screening occurs.
\subsection{Low-lying excitations}
As we discussed above, there is a finite gap between the local level of the boundary bound state and the conduction band. 
At low temperatures, the excitation of an electron from the local level to the Fermi surface must overcome the energy
gap and is thus unimportant to the low temperature thermodynamics. In this subsection, we discuss the low-lying excitations 
near the Fermi surface. Notice that the current excitations in a periodic system\cite{1,2} are forbidden here due to the open 
boundaries\cite{12}.
\par
In the thermodynamic limit, we denote the density functions as $\rho_c(q)$ , $\rho_s(\lambda)$. The energy per unit length
is thus
\begin{eqnarray}
e=\int_{-Q}^Q\epsilon_c^0(q)\rho_c(q)dq,
\end{eqnarray}
where $\epsilon_c^0(q)=2-[q^2+(1/2)^2]^{-1}$ and $\rho_c(q)$ satisfies the following integral equations
\begin{eqnarray}
\rho_c(q)=a_1(q)+\frac1{2L}[\frac1{2\pi} \phi_c'(q)-\delta(q)]-\int_{-Q}^Qa_2(q-q')\rho_c(q')dq'+\int_{-\Lambda}^\Lambda a_1(q-\lambda)\rho_s(\lambda)
d\lambda,\\
\rho_s(\lambda)=\frac1{2L}[\frac1{2\pi}\phi_s'(\lambda)-\delta(\lambda)]+\int_{-Q}^Qa_1(\lambda-q)\rho_c(q)dq-\int_{-\Lambda}^\Lambda a_2(\lambda-
\lambda')\rho_s(\lambda')d\lambda'.
\end{eqnarray}
Here $\phi_{c,s}'$are the derivatives of $\phi_{c,s}^e+\phi_{c,s}^i$; $a_n(q)=\theta_n'(q)/2\pi$. By substituting (50) into (49), we readily obtain that
\begin{eqnarray}
e=\int_{-Q}^Q\epsilon_c(q)a_1(q)dq +\frac 1{4\pi L}[\int_{-Q}^Q\epsilon_c(q)\phi_c'(q)dq+\int_{-\Lambda}^\Lambda
\epsilon_s(\lambda)\phi_s'(\lambda)d\lambda-2\pi \epsilon_c^0(0)],
\end{eqnarray}
where the first term is just the energy of the bulk and the second term is the energy
induced by the impurity and the boundaries;
\begin{eqnarray}
\epsilon_c(q)=\epsilon_c^0(q)-\int_{-Q}^Q a_2(q-q')\epsilon_c(q')dq'+\int_{-\Lambda}^\Lambda a_1(q-\lambda)\epsilon_s(\lambda)
d\lambda,\\
\epsilon_s(\lambda)=\int_{-Q}^Q a_1(\lambda-q)\epsilon_c(q)dq-\int_{-\Lambda}^\Lambda a_2(\lambda-\lambda')\epsilon_s(\lambda')
d\lambda'
\end{eqnarray}
are the dressed energy of the system\cite{29,30}. It seems that the impurity and the open boundary
have no effect to the dressed energies up to the order $O(1/L)$. If the system is at an excited state, some particles in the
lower $I_j$- and $J_\alpha$-states will be moved to higher $I_j$- and $J_\alpha$-states and leave some holes in the Fermi seas.
We denote the momenta of excited particles as $q_p$ and $\lambda_p$, while those of holes as $q_h$ and $\lambda_h$. The bare 
change of the density distributions $\rho_c(q)$ and $\rho_s(\lambda)$ are thus
\begin{eqnarray}
\delta\rho_c^b(q)=\frac1{2L}\sum_{r=\pm}[\sum_p\delta(q-rq_p)-\sum_h\delta(q-rq_h)],\\
\delta\rho_s^b(\lambda)=\frac1{2L}\sum_{r=\pm}[\sum_p\delta(\lambda-r\lambda_p)-\sum_h\delta(\lambda-r\lambda_h)].
\end{eqnarray}
In addition, $\delta\rho_c^b$ and $\delta\rho_s^b$ have feedback effect to $\rho_c$ and $\rho_s$ in the Fermi seas and induce
the change of $\rho_c$ and $\rho_s$
\begin{eqnarray}
\delta\rho_c(q)=\frac1{2L}\sum_{r=\pm}[\sum_h\ a_2(q-rq_h)-\sum_p a_2(q-rq_p)+\sum_p a_1(q-r\lambda_p)
-\sum_h a_1(\lambda-r\lambda_h)]\nonumber\\
-\int_{-Q}^Qa_2(q-q')\delta\rho_c(q')dq'+\int_{-\Lambda}^\Lambda a_1(q-\lambda)\delta\rho_s(\lambda)d\lambda,\\
\delta\rho_s(\lambda)=\frac1{2L}\sum_{r=\pm}[\sum_pa_1(\lambda-rq_p)- \sum_h a_1(\lambda-rq_h)-
\sum_pa_2(\lambda-r\lambda_p)+\sum_ha-2(\lambda-
r\lambda_h)]\nonumber\\
+\int_{-Q}^Qa_1(\lambda-q)\delta\rho_c(q)dq-\int_{-\Lambda}^\Lambda a_2(\lambda-\lambda')\delta\rho_s(\lambda')d\lambda'.
\end{eqnarray}
The excitation energy is
\begin{eqnarray}
\Delta E=\sum_p\epsilon_c^0(q_p)-\sum_h\epsilon_c^0(q_h)+\int_{-Q}^Q\epsilon_c^0(q)\delta\rho_c(q)\nonumber\\
=\sum_p\epsilon_c(q_p)-\sum_h\epsilon_c(q_h)+\sum_p\epsilon_s(\lambda_p)-\sum_h\epsilon_s(\lambda_h).
\end{eqnarray}
It is just the same expression for those of a pure system\cite{29,30}. That means the impurity does not affect the bulk excitations
up to the order $O(1/L)$. The dressed energies $\epsilon_{c,s}$ thus can be treated as quasi-particle energies in the so-called
Landau-Luttinger liquid theory\cite{31}. Nevertheless, the impurity indeed affects the thermodynamics in the order of $
O(1/L)$ since it induces the changes of the density of states in this order as shown in (50), (51). 
We remark there are also string
excitations in the spin sector\cite{28}. The excitation energy can be expressed as the same form of (59) and the strings induce only
different scattering matrices from those of spin holes\cite{28}.
\section{Thermodynamics}

In this section, we derive the thermodynamic equations of the present model 
via thermal Bethe ansatz\cite{29,32}. We shall
omit the excitations which break the boundary bound state since the excitations are accompanied by finite energy gaps 
and their
contributions to the low temperature thermodynamic quantities are exponentially small. 
\subsection{Thermodynamic Bethe ansatz}
At finite temperatures, the solution of 
the Bethe ansatz equations are described by a sequence of real $\{q_j\}$ and a variety of $\{\lambda_\alpha\}$ strings. From
the Bethe ansatz equations we obtain
\begin{eqnarray}
\rho_c(q)+\rho_c^h(q)=a_1(q)+\frac1{4\pi L}\phi_c'(q)-\frac1{2L}\delta(q)-[2]\rho_c(q)+\sum_{m=1}^\infty [m]\rho_{s,m}(q),\\
\rho_{s,m}^h(\lambda)=\frac 1{4\pi L}\phi_{s,m}'(\lambda)-\frac1{2L}\delta(\lambda)+
[m]\rho_c(\lambda)-\sum_{n=1}^\infty A_{mn}\rho_{s,n}(\lambda),
\end{eqnarray}
where $\rho_{s,m}$ $(\rho_{s,m}^h)$ are the densities of the $m$-string (holes);$[n]$ is an integral operator with the
kernel $a_n(q)$; $A_{mn}=[m+n]+2[m+n-2]+\cdots+[|m-n|+1]$;
$\phi_{s,m}=\sum_{j=1}^m\phi_s[\lambda+i((m+1)/2-j)]$.  In a magnetic field $H$, the free energy
of the system can be written as
\begin{eqnarray}
F/L=\int[\epsilon_c^0(q)-\mu-\frac H2]\rho_c(q)dq+\sum_{n=1}^\infty nH\int\rho_{s,n}(\lambda) d\lambda\nonumber\\
-T\int[(\rho_c+\rho_c^h)\ln(\rho_c+\rho_c^h)-\rho_c\ln\rho_c-\rho_c^h\ln\rho_c^h]dq\nonumber\\
-T\sum_{n=1}^\infty\int[(\rho_{s,n}+\rho_{s,n}^h)\ln(\rho_{s,n}+\rho_{s,n}^h)-\rho_{s,n}\ln\rho_{s,n}-\rho_{s,n}^h\ln
\rho_{s,n}^h]d\lambda.
\end{eqnarray}
At equilibrium state, the free energy must be minimum. This gives
\begin{eqnarray}
\ln\eta=\frac{\epsilon_c^0-\mu-\frac12H}T+[2]\ln(1+\eta^{-1})-\sum_n[n]\ln(1+\zeta_n^{-1}),\\
\ln(1+\zeta_n)=\frac{nH}T-[n]\ln(1+\eta^{-1})+\sum_m A_{mn}\ln(1+\zeta_m^{-1}),
\end{eqnarray}
where $\eta(q)=\rho_c^h(q)/\rho_c(q)$, $\zeta_n(\lambda)=\rho_{s,n}^h(\lambda)/\rho_{s,n}(\lambda)$. The above equations
can be rewritten as
\begin{eqnarray}
\ln\eta=\frac{\epsilon_0-\mu}T+\{[2]-[1]G\}\ln(1+\eta^{-1})-G\ln(1+\zeta_1),\\
\ln\zeta_n=G[\ln(1+\zeta_{n-1})+\ln(1+\zeta_{n+1})],{~~~~}n>1,\\
\ln\zeta_1=-G\ln(1+\eta^{-1})+G\ln(1+\zeta_2),
\end{eqnarray}
where $G$ is an integral operator with the kernel $[2\cosh(\pi\lambda)]^{-1}$. The free energy takes the form
\begin{eqnarray}
F=F_0+\frac1L F_{imp}+\frac1L F_{edg},\\
F_0/L=-T\int a_1(q)\ln(1+\eta^{-1}(q)]dq,\\
F_{edg}=\frac12T\ln(1+\eta^{-1}(0)]+\frac12T\sum_n\ln[1+\zeta_n^{-1}(0)]\nonumber\\
-\frac T{4\pi}\int {\phi_c^e}'(q)\ln(1+\eta^{-1})dq-\frac T{4\pi}\sum_n\int{\phi_{s,n}^e}'(\lambda)
\ln(1+\zeta_n^{-1})d\lambda,\\
F_{imp}=-\frac T{4\pi}\int {\phi_c^i}'(q)\ln(1+\eta^{-1})dq-\frac T{4\pi}\sum_n\int{\phi_{s,n}^i}'(\lambda)
\ln(1+\zeta_n^{-1})d\lambda,
\end{eqnarray}
where $F_0$ is the free energy of the bulk, $F_{edg}$ is the free energy of the open boundary and $F_{imp}$ is the free 
energy of the impurity which contains a part of
the boundary energy\cite{24}, the charge-fluctuation term and the contribution of the impurity spin.
\subsection{Residual Entropy}
To give a detailed picture of the ground state configuration, we study the residual entropy of the impurity. When $T\to 0$,
$\eta\to\exp[(\epsilon_c-\mu)/T],{~~} for{~~} \epsilon_c<\mu $. Therefore the driving term in (67) tends to $-\infty$. That means
$\zeta_1\to 0$ and all the other $\zeta_n$ take constant values\cite{28,32} $\zeta_n^+$ with
\begin{eqnarray}
\zeta_n^+=\frac{\sinh^2(nx_0)}{\sinh^2 x_0}-1,{~~~~}x_0=\frac H{2T}.
\end{eqnarray}
Here we consider the spin part of the impurity free energy $F_{imp}^s$, which is the only relevant term
to the residual entropy and the magnetization. With different $c$, the ground state configurations are very different. Below
we derive the residual entropy for different parameter regions. Since at zero temperature, $\zeta_n$ are variable-independent,
the integral operators $[n]$ are equivalent to unit and $a_n$ are equivalent to  $\delta (\lambda)$. This makes the calculations
very simple. For half-integer or integer $c$, the impurity spin is split into two effective spins $S_1$ and $S_2$ and the
residual entropy reads
\begin{eqnarray}
S_{res}=\frac12sgn(S_1)\ln(2|S_1|)+\frac12sgn(S_2)\ln(2|S_2|).
\end{eqnarray}
The following discussion is for non-half-integer $c$ case.
\par
(i)$c > S+1$. In this case,
\begin{eqnarray}
\frac1{2\pi}{\phi_{s,n}^i}'(\lambda)\to a_{n,2c_I+2S-1}(\lambda)-a_{n,2c_I-2S-1}(\lambda)-a_\alpha(\lambda)
\sum_{l=1}^{2S}\delta_{n,2c_I-2S+2l-2}, 
\end{eqnarray}
where $c_I$ is defined as: if $n/2\leq|c|<(n+1)/2$, $c_I=|n|/2$; $\alpha=2(|c|-c_I)$; and
\begin{eqnarray}
a_{n,m}=\sum_{k=1}^{min(m,n)}a_{n+1+m-2k}.
\end{eqnarray}
For $T\to 0$,
\begin{eqnarray}
F_{imp}^s\to-\frac14 T\ln\frac{1+\zeta_{2c_I+2S-1}^+}{1+\zeta_{2c_I-2S-1}^+}+\frac12T\sum_{l=1}^{2S}\ln(1+
\zeta{_{2c_I-2S+2l-2}^+}^{-1}).
\end{eqnarray}
Therefore, the residual entropy takes the value
\begin{eqnarray}
S_{res}=\frac12\ln\frac{2c_I+2S-1}{2c_I-2S-1}-\sum_{l=1}^{2S}\ln\frac{2(c_I-S-1+l)}{\sqrt{4(c_I-S-1+l)^2-1}}.
\end{eqnarray}
The magnetization at zero temperature is $M=S$, which is the same result given in sec. III.
\par
(ii)$S+1>c>S+1/2$. In this case, a boundary bound state exists in the ground state and $c_I=S+1/2$.
\begin{eqnarray}
\frac1{2\pi}{\phi_{s,n}^i}'(\lambda)\to a_{n,4S}+a_{n,2}-a_\alpha\sum_{l=1}^{2S-1}\delta_{n,2l+1}.
\end{eqnarray}
The residual entropy reads
\begin{eqnarray}
S_{res}=\ln(2\sqrt{2S})-\sum_{l=1}^{2S-1}\ln\frac{2l+1}{\sqrt{4l(l+1)}}.
\end{eqnarray}
\par
(iii)$1/2<c<S+1/2$. No bound state exists and
\begin{eqnarray}
\frac1{2\pi}{\phi_{s,n}^i}'(\lambda)\to a_{n,2S+2c_I-1}+a_{n,2S-2c_I+1}-a_\alpha\sum_{l=1}^{2c_I-1}\delta_{n,2S-2c_I+2l}.
\end{eqnarray}
This gives
\begin{eqnarray}
S_{res}=\ln\sqrt{4S^2-(2c_I-1)^2}-\sum_{l=1}^{2c_I-1}\ln\frac{2S-2c_I+2l}{\sqrt{(2S-2c_I+2l)^2-1}}.
\end{eqnarray}
\par
(iv)$c_I=0$. In this case, $c$ has no effect to the residual entropy and
\begin{eqnarray}
\frac1{2\pi}{\phi_{s,n}^i}'(\lambda)\to 2a_{n,2S}.
\end{eqnarray}
The residual entropy reads
\begin{eqnarray}
S_{res}=ln{(2S)},
\end{eqnarray}
a similar result to that in the conventional Kondo problem\cite{28}.
\par
(v)$-(S-1/2)<c<-1/2$.
\begin{eqnarray}
\frac1{2\pi}{\phi_{s,n}^i}'(\lambda)\to a_{n,2S+1+2c_I}+a_{n,2S-1-2c_I}-a_\alpha\sum_{l=0}^{2c_I}\delta_{n,2S-2c_I+2l},
\end{eqnarray}
and
\begin{eqnarray}
S_{res}=\ln\sqrt{4S^2-(2c_I+1)^2}-\sum_{l=0}^{2c_I}\ln\frac{2S-2c_I+2l}{\sqrt{(2S-2c_I+2l)^2-1}}.
\end{eqnarray}
\par
(vi)$-S<c<-(S-1/2)$. In this case, $c_I=S-1/2$ and
\begin{eqnarray}
\frac1{2\pi}{\phi_{s,n}^i}'(\lambda)\to a_{n,4S}-a_\alpha\sum_{l=0}^{2S-1}\delta_{n,2l+1}.
\end{eqnarray}
Caution should be taken in this case since $\ln(1+\zeta_1^{-1})$ appears in the free energy, which is divergent when 
$T\to 0$. Note that $2\ln\zeta_1\to-\ln(1+\eta^{-1})+\ln(1+\zeta_2)$ as shown in (67) and $\ln(1+\eta^{-1})$ takes the 
asymptotic
form $AT^{-1}+BT$. Therefore $\eta$ contributes nothing to the entropy. The residual entropy in this case reads
\begin{eqnarray}
S_{res}=\ln(2\sqrt{2S})-\sum_{l=1}^{2S-1}\ln\frac{2l+1}{2\sqrt{l^2+l}}.
\end{eqnarray}
\par
(vii)$-(S+1/2)<c<-S$. In this case, boundary bound states exist in the ground state and $c_I=S$.
\begin{eqnarray}
\frac1{2\pi}{\phi_{s,n}^i}'(\lambda)\to a_{n,4S+1}-a_{n,3}-a_\alpha\sum_{l=1}^{2S-1}\delta_{n,2+2l}.
\end{eqnarray}
\begin{eqnarray}
S_{res}=\frac12\ln\frac{4S+1}3-\sum_{l=1}^{2S-1}\ln\frac{2(l+1)}{\sqrt{(4(l+1)^2-1}}.
\end{eqnarray}
\par
(viii)$c<-(S+1/2)$. The situation of this case is very similar to that of case (i) and
\begin{eqnarray}
\frac1{2\pi}{\phi_{s,n}^i}'(\lambda)\to a_{n,2S+2c_I+1}-a_{n,2c_I-2S+1}-a_\alpha\sum_{l=1}^{2S}\delta_{n,2c_I-2S+2l}.
\end{eqnarray}
The residual entropy reads
\begin{eqnarray}
S_{res}=\ln\sqrt{(2c_I+1)^2-4S^2}-\sum_{l=1}^{2S}\ln\frac{2(c_I-S+l)}{\sqrt{4(c_I-S+l)^2-1}}.
\end{eqnarray}
From the above discussion we can see that the impurity-bulk coupling not only changes the effective energy scale as in the
conventional Kondo problem, but also renormalize the effective strength of the impurity spin. Generally, the coupling splits
the impurity spin into two effective spins (ghost spins). At low temperatures, the local spin configuration is strongly
interaction-dependent rather than simply screened or decoupled as understood in the conventional Kondo problem. This strongly
suggests that the correlation  in the bulk has a non-trivial effect to the low-temperature behavior of the impurity.
\subsection{Local Landau-Luttinger liquid description}
Despite the finite residual entropy in the ground state, the leading order of the low temperature specific heat should be
T-linear since there is no extra degrees of freedom to induce the overscreening effect. Therefore, 
Nozi\`{e}res' Local Fermi
liquid theory\cite{33} can be used with a slight modification. Consider a non-interacting 1D open chain. The ``momenta" $k'$
is quantized as
\begin{eqnarray}
k'=k+\frac1{2L}\delta_0(k),
\end{eqnarray}
where $k=\pi n/L$ ($n$ positive integer) are the ``momenta" of the pure open boundary system and $\delta_0(k)$ is the phase
shift due to the particle-impurity scattering. The change of the density of states is therefore
\begin{eqnarray}
\delta D(k)=\frac1{2\pi}\frac{\delta_0'(k)}{\epsilon'(k)},
\end{eqnarray}
where $\epsilon(k)$ is the single-particle energy. For the interacting 1D electron systems, the quasi-particles can be defined
in the charge- and spin-sectors respectively\cite{31}. The phase shifts are generally functionals of the distribution of
the quasi-particles $\delta_c[n_c(q), n_s(\lambda)]$, $\delta_s[n_c(q),n_s(\lambda)]$ and can be expressed as
\begin{eqnarray}
\delta_c(q)=\delta_{c,0}(q)+\sum_{r=\pm}\sum_{q'\neq q}\theta_{cc}(q,rq')\delta n_c(q')+\sum_{r=\pm}\sum_{\lambda}\theta_{cs}(q,r\lambda)
\delta n_s(\lambda),\\
\delta_s(\lambda)=\delta_{s,0}(\lambda)+\sum_{r=\pm}\sum_{q}\theta_{sc}(\lambda,rq)\delta n_c(q)+\sum_{r=\pm}\sum_{\lambda'
\neq \lambda}\theta_{ss}(\lambda,r\lambda')\delta n_s(\lambda'),
\end{eqnarray}
where $\delta_{c,0}$ and $\delta_{s,0}$ are the bare phase shifts induced by the impurity; $\theta$'s are the phase shifts
of the particle-particle scattering; $\delta n_{c,s}=n_{c,s}-n_{c,s}^0$ is the change of the quasi-particle distributions
induced by the impurity. In another hand, $\delta n_{c,s}$ in the ground state take the form
\begin{eqnarray}
\delta n_c(q)=\frac{\delta_c'(q)}{2 L},{~~~~~}\delta n_c(\lambda)=\frac{\delta_s'(\lambda)}{2 L}.
\end{eqnarray}
The equations (94) and (95) are thus reduced in the thermodynamic limit $L\to \infty$ to
\begin{eqnarray}
\delta_c'(q)=\delta_{c,0}'(q)+\frac1{2\pi}\int_{-q_F}^{q_F}\theta_{cc}'(q,q')\delta_c'(q')dq'+\frac1{2\pi}\int_{-\lambda_F}
^{\lambda_F}\theta_{cs}'(q,\lambda)\delta_s'(\lambda)d\lambda,\\
\delta_s'(\lambda)=\delta_{s,0}'(\lambda)+\frac 1{2\pi}\int_{-q_F}^{q_F}\theta_{sc}(\lambda,q)\delta_c'(q)dq+\frac 1{2\pi}
\int_{-\lambda_F}^{\lambda_F}\theta_{ss}'(\lambda,\lambda')\delta_s'(\lambda')d\lambda',
\end{eqnarray}
where $q_F$ and $\lambda_F$ are the Fermi ``momenta". Notice $o(1/L)$ terms have been omitted in the above equations. The
low-temperature thermodynamics of the impurity is therefore charactered by two constants
\begin{eqnarray}
\delta D_c(q_F)=\frac1{2\pi}\frac{\delta'(q_F)}{\epsilon_c'(q_F)},{~~~~}\delta D_s(\lambda_F)=\frac 1{2\pi}\frac{\delta_s'(
\lambda_F)}{\epsilon'_s(\lambda_F)},
\end{eqnarray}
where $\epsilon_{c,s}$ are the quasi-particle energies.
\par
In our case,
\begin{eqnarray}
\delta_{c,0}(q)=\phi_c(q),{~~~~}\delta_{s,0}(\lambda)=\phi_s(\lambda),\\
\theta_{cc}'(q,q')=-2\pi a_2(q-q'),{~~~~}\theta_{cs}'(q,\lambda)=2\pi a_1(q-\lambda),\\
\theta_{sc}'(\lambda,q)=2\pi a_1(\lambda-q),{~~~~}\theta_{ss}'(\lambda-\lambda')=-2\pi a_2(\lambda-\lambda').
\end{eqnarray}
Therefore,
\begin{eqnarray}
\delta_c'(q)=\phi_c'(q)-\int_{-Q}^Q a_2(q-q')\delta_c'(q')dq'+\int_{-\Lambda}^\Lambda a_1(q-\lambda)\delta_s'(\lambda)d\lambda,\\
\delta_s(\lambda)=\phi_s'(\lambda)+\int_{-Q}^Q a_1(\lambda-q)\delta_c'(q)dq-\int_{-\Lambda}^\Lambda a_2(\lambda-\lambda')\delta_s'(\lambda')d\lambda'.
\end{eqnarray}
The variations of the densities of states at the Fermi surfaces induced by the impurity are given by (99) with 
$\epsilon_{c,s}$
the dressed energies (53-54). For the low-temperature specific heat, the following relation holds
\begin{eqnarray}
\frac{\delta C}{C_0}=\frac{\delta_c'(Q)}{\rho_c(Q)}+\frac{\delta_s'(\Lambda)}{\rho_s(\Lambda)},
\end{eqnarray}
where $\delta C$ is the specific heat induced by the impurity and the open boundaries and $C_0$ is the specific heat of the bulk
(per unit length); $\rho_c(q)$ and $\rho_s(\lambda)$ are the quasi-particle distributions in the ground state.
\section{concluding remarks}
In conclusion, we propose an integrable model of a boundary impurity spin  coupled with the integrable  open $t-J$ chains.
In our
model,  The ``fine tuned" effect in the periodic integrable models is overcome and the interaction term 
takes a very simple form. The coupling constant $J$ can take arbitrary value without destroying the integrability of
the Hamiltonian, while in the periodic models, there is a constraint to $J$. Though a similar impurity
can be introduced in the periodic models, the parameter $c$ must be real (imaginary in our case) which describes a
weak-linked impurity to the bulk. The interaction only affects the energy scale (Kondo temperature) but does not change
the fixed point of the system. With an imaginary $c$, the model Hamiltonians constructed for bulk impurities are non-Hermitian
and their spectra generally lie in the complex plane rather than in the real axis completely. In our model, 
real $c$  defines a Hermitian Hamiltonian due to the reflection symmetry, and the coupling constant $J$ meets
all physical situations. Some new  phenomena driven by the impurity-bulk coupling have been found, which can
never appear in the periodic models as well as in the conventional Kondo problems:(i)The stronger coupling $J$ may 
split the impurity spin into effective ``ghost spins"
$S-c_I+1/2$ and $S+c_I-1/2$. That means the coupling not only changes the energy scales (Kondo temperature) as in the conventional
Kondo problem but also renormalizes the effective strength of the impurity spin. It seems that the strength of these
ghost spins does not change via temperature. Such a phenomenon reveals a pure correlation effect. 
(ii)Depending on the strength of the coupling,
the system may show  behavior differing from those of the conventional Kondo problems. A typical example
is that the scalar potential may destruct the Kondo screening even in the antiferromagnetic regime. (iii)The residual entropy
is strongly coupling-dependent which hints the local spin configuration near the impurity is very complicated rather than 
simply
screened or decoupled as understood in the conventional Kondo problem.
\par
One of the authors (YW) acknowledges the financial aids of AvH-Stiftung and CNFNS.

\begin{center}
{\bf Appendix A}
\end{center}
\par
Since the model is spin $SU(2)$ invariant, the reflection matrix must take the following form
$$
R_j(q_j)=f(q_j)[g(q_j)+i\sigma_j\cdot{\bf S}],\eqno{(A.1)}
$$
where $f(q)$ is an arbitrary function and $g(q)$ is determined by the reflection equation. With an $S-$matrix
$$
S_{jl}(q_j,q_l)=\frac{q_j-q_l+i\gamma P_{jl}}{q_j-q_l+i\gamma},\eqno{(A.2)}
$$
from the reflection equation (3) we have
$$
[q_j-q_l+i\gamma P_{jl}][g(q_j)+i\sigma_j\cdot{\bf S}][q_j+q_l+i\gamma P_{jl}][g(q_l)+i\sigma_l
\cdot{\bf S}]
$$
$$
=[g(q_l)+i\sigma_l\cdot{\bf S}][q_j+q_l+i\gamma P_{jl}][g(q_j)+i\sigma_j\cdot{\bf S}][q_j-q_l+
i\gamma P_{jl}]. \eqno{(A.3)}
$$
By virtue of the relation $(\sigma\cdot{\bf S})^2=S(S+1)-\sigma\cdot{\bf S}$, we obtain
$$
\gamma q_j[2g(q_j)-i]-q_j^2=\gamma q_l[2g(q_l)-i]-q_l^2.\eqno{(A.4)}
$$
Since $q_j,q_l$ are arbitrary, the left hand side and the right hand side of the above equation must be a constant $C$. 
Therefore,
$$
g(q)=\frac q{\gamma}+\frac C{\gamma q}+\frac i2.\eqno{(A.5)}
$$
Substituting (A.5) into (A.1), we readily obtain
$$
R_j(q_j)=e^{i\varphi(q_j)}\frac{q_j-ic\gamma+i\gamma(\sigma_j\cdot{\bf S}+1)}{q_j-ic\gamma+i\gamma(S+1)}\frac{q_j+ic\gamma
+i\gamma\sigma_j\cdot{\bf S}}{q_j+ic\gamma+i\gamma S},\eqno{(A.6)}
$$
where $\varphi(q)$ is an arbitrary function and $c$ is an arbitrary constant.
\par
\begin{center}
{\bf Appendix B}
\end{center}
In dealing with the open boundary integrable models, we often encounter the following eigenvalue problem
$$
S_{jj-1}^-\cdots S_{j1}^-S_{j1}^+\cdots S_{jj-1}^+S_{jj+1}^+\cdots S_{jN}^+R_jS_{jN}^-\cdots S_{jj+1}^-\psi_0\equiv X_j\Psi_0
=\epsilon(q_j)\psi_0,\eqno{(B.1)}
$$
where $S_{jl}^\pm$ and $R_j$ are the scattering matrix and the reflection matrix respectively. In our case, up to constants they take
the forms
$$
S_{jl}^\pm=\frac{q_j\pm q_l+i\gamma P_{jl}}{q_j\pm q_l+i\gamma},\eqno{(B.2)}
$$
$$
R_j=\frac{q_j-ic\gamma+i\gamma(\sigma_j\cdot{\bf S}+1)}{q_j-ic\gamma+i\gamma(S+1)}\frac{q_j+ic\gamma
+i\gamma\sigma_j\cdot{\bf S}}{q_j+ic\gamma+i\gamma S}.\eqno{(B.3)}
$$
Define ${\bar\psi}_0=(S_{jj-1}^-\cdots S_{j1}^-)^{-1}\psi_0$. $(B.1)$ can be rewritten as
$$
S_{j1}^+\cdots S_{jj-1}^+S_{jj+1}^+\cdots S_{jN}^+R_jS_{jN}^-\cdots S_{jj+1}^-S_{jj-1}^-\cdots S_{j1}^-{\bar\psi}_0
\equiv X_j'{\bar\psi}_0=\epsilon(q_j){\bar \psi}_0.\eqno{(B.4)}
$$
For convenience, we introduce an auxiliary space $\tau$ and define
$$
U_\tau(q)=S_{\tau j}^+S_{\tau1}^+\cdots S_{\tau j-1}^+\cdots S_{\tau N}^+R_{\tau0}S_{\tau N}^-\cdots S_{\tau j+1}^-
S_{\tau j-1}^-\cdots S_{\tau 1}^-S_{\tau j}^-,\eqno{(B.5)}
$$
with $q_\tau=q$. Obviously, $S_{\tau j}^-(q_j)=P_{\tau j}$ and
$$
tr_\tau U_\tau(q_j)=\frac{2q_j+2i\gamma}{2q_j+i\gamma}X_j'.\eqno{(B.6)}
$$
Since $S_{\tau l}^\pm$ satisfy the Yang-Baxter relation
$$
S_{\tau\tau'}^-(q-q')S_{\tau j}^\pm(q\pm q_j)S_{\tau' j}^\pm(q'\pm q_j)= S_{\tau' j}^\pm(q'\pm q_j)S_{\tau j}^\pm(q\pm q_j)
S_{\tau\tau'}^-(q-q'),\eqno{(B.7)}
$$
from (3) we can easily show that $U_\tau (q)$ satisfies the reflection equation
$$
S_{\tau\tau'}^-(q-q')U_\tau (q)S_{\tau\tau'}^+(q+q')U_{\tau'}(q')=U_{\tau'}(q')S_{\tau\tau'}^+(q+q')U_\tau (q)
S_{\tau\tau'}^-(q-q').\eqno{(B.8)}
$$
Therefore, the eigenvalue problem (B.3) is reduced to Sklyanin's eigenvalue problem\cite{18}. Furthermore, we define
$$
T_\tau(q)=S_{\tau_0}(q)S_{\tau N}^-(q)\cdots S_{\tau j+1}^-\cdots S_{\tau j-1}^-\cdots S_{\tau1}^-S_{\tau j}^-(q),\eqno{(B.9)}
$$
where
$$
S_{\tau 0}^-=\frac{q+i\gamma c+i\gamma \tau\cdot{\bf S}}{q+i\gamma c+i\gamma S}.\eqno{(B.10)}
$$
Then 
$$
U_\tau(q)=T_\tau^{-1}(-q)T_{\tau}(q),\eqno{(B.11)}
$$
and $T_\tau(q)$ satisfies
$$
S_{\tau\tau'}^-(q-q')T_\tau(q)T_{\tau'}(q')=T_{\tau'}(q')T_\tau(q)S_{\tau\tau'}^-(q-q').\eqno{(B.12)}
$$
For convenience, we introduce the notations
\begin{eqnarray*}
U_\tau(q)=\pmatrix{A(q),&B(q) \cr C(q),&D(q)},
{~~~}T_\tau(q)= \pmatrix{a(q),&b(q) \cr c(q),&d(q)}.
\end{eqnarray*}

Notice that 
$$
\tau_2{S_{\tau j}^-}^t(-q-i\gamma)\tau_2=S_{\tau j}^+(q)(1+\frac{i\gamma}{q+q_j}),
$$
$$
\tau_2{S_{\tau 0}^-}^t(-q-i\gamma)\tau_2=S_{\tau 0}^+(q)[\frac{q-ic\gamma+i\gamma(S+1)}
{q-ic\gamma-i\gamma(S+1)}].\eqno{(B.13)}
$$
Therefore
$$
T_\tau^{-1}(-q)=\frac{q-ic\gamma-i\gamma(S+1)}{q-ic\gamma+i\gamma(S+1)}\prod_{j=1}^N\frac{q+q_j}{q+q_j+i\gamma}\tau_2
T_\tau(-q-i\gamma)\tau_2.\eqno{(B.14)}
$$
From (B.7) and (B.12) we have the following commutation relations
$$
[c(q),b(q')]=-\frac{i\gamma}{q-q'}\{d(q')a(q)-d(q)a(q')\},\eqno{(B.15)}
$$
$$
[A(q)+D(q),A(q')+D(q')]=0,\eqno{(B.16)}
$$
$$
A(q)B(p)=\frac{(q+p)(q-p-i\gamma)}{(q-p)(q+p+i\gamma)}B(p)A(q)-\frac{i\gamma}{(q+p+i\gamma)(2p+i\gamma)}B(p){\bar D}(p)
$$
$$
+\frac{2i\gamma p}{(q-p)(2p+i\gamma)}B(p)A(q),\eqno{(B.17)}
$$
$$
{\bar D}(q)B(p)=\frac{(q-p+i\gamma)(q+p+2i\gamma)}{q-p)(q+p+i\gamma)}B(p){\bar D}(q)-\frac{2i\gamma(q+i\gamma)}
{(q-p)(2p+i\gamma)}B(q){\bar D}(p)
$$
$$
+\frac{4i\gamma(q+i\gamma)p}{(2p+i\gamma)(q+p+i\gamma)}B(q)A(p),\eqno{(B.18)}
$$
where ${\bar D}(q)=(2q+i\gamma)D(q)-i\gamma A(q)$. To construct the eigenstates of $X_j'$, we define the pseudovacuum 
$|\Omega>$
as all spin up. From the definition we know
$$
a(q)|\Omega>=|\Omega>,{~~~~}c(q)|\Omega>=0,
$$
$$
d(q)|\Omega>=\frac{q+i\gamma c-i\gamma S}{q+i\gamma c+i\gamma S}\prod_{j=1}^N\frac{q-q_j}{q-q_j+i\gamma}|\Omega>.\eqno{(B.19)}
$$
Therefore
$$
A(q)|\Omega>=|\Omega>,{~~~~}C(q)|\Omega>=0.\eqno{(B.20)}
$$
Notice that
$$
D(q)=[a(-q-i\gamma)d(q)-c(-q-i\gamma)b(q)]\frac{q-ic\gamma-i\gamma(S+1)}{q-ic\gamma+i\gamma(S+1)}
\prod_{j=1}^N\frac{q+q_j}{q+q_j+i\gamma}.\eqno{(B.21)}
$$
With the commutation relations (B.15) and (B.19), we obtain
$$
{\bar D}(q)|\Omega>=2q\frac{q-i\gamma(S+c-1)}{q+i\gamma(S-c+1)}\frac{q-i\gamma(S-c)}{q+i\gamma(S+c)}\prod_{j=1}^N
\frac{(q+q_j)(q-q_j)}{(q+q_j+i\gamma)(q-q_j+i\gamma)}|\Omega>.\eqno{(B.23)}
$$
The operator $B(q)$ can be treated as the creation operator of the eigenstate. An eigenstate of $X'(q)$ with $M$ spin flipped
can be written as $\prod_{\alpha=1}^M B(\Lambda_\alpha)|\Omega>\equiv Q_M|\Omega>$. From $(B.17)$ and
$(B.18)$ we have the following relations
$$
A(q)Q_M=Q_MA(q)\prod_{\alpha=1}^M\frac{(q+\Lambda_\alpha)(q-\Lambda_\alpha-i\gamma)}{(q-\Lambda_\alpha)(q+\Lambda_\alpha
+i\gamma)}
$$
$$
-\sum_{\alpha=1}^M\frac{i\gamma}{(q+\Lambda_\alpha+i\gamma)(2\Lambda_\alpha+i\gamma)}Q_M^\alpha{\bar
D}(\Lambda_\alpha)\prod_{\beta\neq \alpha}\frac{(\Lambda_\alpha-\Lambda_\beta+i\gamma)(\Lambda_\alpha+
\Lambda_\beta+2i\gamma)}{(\Lambda_\alpha-\Lambda_\beta)(\Lambda_\alpha+\Lambda_\beta+i\gamma)}
$$
$$
+\sum_{\alpha=1}^M\frac{2i\gamma\Lambda_\alpha}{(q-\Lambda_\alpha)(2\Lambda_\alpha+i\gamma)}Q_M^\alpha
A(\Lambda_\alpha)\prod_{\beta\neq \alpha}\frac{(\Lambda_\alpha-\Lambda_\beta)(\Lambda_\alpha+
\Lambda_\beta-i\gamma)}{(\Lambda_\alpha-\Lambda_\beta)(\Lambda_\alpha+\Lambda_\beta+i\gamma)},\eqno{(B.24)}
$$
$$
{\bar D}(q)Q_M=Q_M{\bar D}(q)\prod_{\alpha=1}^M\frac{(q+\Lambda_\alpha+i\gamma)(q-\Lambda_\alpha+
2i\gamma)}{(q-\Lambda_\alpha)(q+\Lambda_\alpha
+i\gamma)}
$$
$$
-\sum_{\alpha=1}^M\frac{2i\gamma(q+i\gamma)}{(q-\Lambda_\alpha)(2\Lambda_\alpha+i\gamma)}Q_M^\alpha{\bar
D}(\Lambda_\alpha)\prod_{\beta\neq \alpha}\frac{(\Lambda_\alpha-\Lambda_\beta+i\gamma)(\Lambda_\alpha+
\Lambda_\beta+2i\gamma)}{(\Lambda_\alpha-\Lambda_\beta)(\Lambda_\alpha+\Lambda_\beta+i\gamma)}
$$
$$
+\sum_{\alpha=1}^M\frac{4i\gamma\Lambda_\alpha(q+i\gamma)}{(q+\Lambda_\alpha+i\gamma)
(2\Lambda_\alpha+i\gamma)}Q_M^\alpha
A(\Lambda_\alpha)\prod_{\beta\neq \alpha}\frac{(\Lambda_\alpha-\Lambda_\beta)(\Lambda_\alpha+
\Lambda_\beta-i\gamma)}{(\Lambda_\alpha-\Lambda_\beta)(\Lambda_\alpha+\Lambda_\beta+i\gamma)},\eqno{(B.25)}
$$
where $Q_M^\alpha=B(\Lambda_1)\cdots B(\Lambda_{\alpha-1})B(q)B(\Lambda_{\alpha+1})\cdots B(\Lambda_M)$. Obviously,
when we apply $A(q)$ and ${\bar D}(q)$ to the state $Q_M|\Omega>$, two types of terms appear. One is the wanted
terms with $Q_M$, the other is unwanted terms with $Q_M^\alpha$. With the relation $X_j'(q_j)=[
2(q_j+i\gamma)]^{-1}{\bar D}(q_j)
+A(q_j)$, we obtain
$$
\epsilon(q_j)=\prod_{\alpha=1}^M\frac{q_j+\Lambda_\alpha}{q_j-\Lambda_\alpha}\frac{q_j-\Lambda_\alpha-i\gamma}
{q_j+\Lambda_\alpha+i\gamma}.\eqno{(B.26)}
$$
The cancellation of the unwanted terms gives the constriction of the parameters $\Lambda_\alpha$
$$
\frac{\Lambda_\alpha-i\gamma(S+c-1)}{\Lambda_\alpha+i\gamma(S+c)}\frac{\Lambda_\alpha-i\gamma(S-c)}{\Lambda_\alpha+i\gamma(S-c+1)}
\prod_{j=1}^N\frac{(\Lambda_\alpha+q_j)(\Lambda_\alpha-q_j)}{(\Lambda_\alpha+q_j+i\gamma)
(\Lambda_\alpha-q_j+i\gamma)}
$$
$$
=\prod_{\beta\neq \alpha}\frac{(\Lambda_\alpha+\Lambda_\beta)(\Lambda_\alpha-\Lambda_\beta-i\gamma)}
{(\Lambda_\alpha+\Lambda_\beta+2i\gamma)(\Lambda_\alpha-\Lambda_\beta+i\gamma)}.\eqno{(B.27)}
$$
By replacing $\Lambda_\alpha$ with $\lambda_\alpha-i\gamma/2$, we readily obtain the Bethe ansatz equations.


\begin{references}
\bibitem{1}F.D.M. Haldane, J. Phys. C {\bf 14}, 2585(1981).
\bibitem{2}J. Voit, Rep. Prog. Phys. {\bf 58}, 977(1995).
\bibitem{3}P.W. Anderson, Phys. Rev. Lett. {\bf 64}, 1839(1990); {\bf 65}, 2306(1991).
\bibitem{4}A.O. Gogolin, Ann. Phys. (Paris), {\bf 19}, 411(1994).
\bibitem{5}K. Moon, C.L. Kane, S. M. Girvin and M.P.A. Fisher, Phys. Rev. Lett. {\bf 71}, 4381(1993).
\bibitem{6}C.L. Kane and M.P.A. Fisher, Phys. Rev. Lett. {\bf 68}, 1220(1992); Phys. Rev. B {\bf 46}, 15233(1992).
\bibitem{7}S. Eggert and I. Affleck, Phys. Rev. B {\bf 46}, 10866(1992).
\bibitem{8}I. Affleck and W.W. Ludwig, J. Phys. A {\bf 27}, 5375(1994).
\bibitem{9}D.H. Lee and J. Toner, Phys. Rev. Lett. {\bf 69}, 3378(1992).
\bibitem{10}A. Furusaki and N. Nagaosa, Phys. Rev. Lett. {\bf 72}, 892(1994).
\bibitem{11}P. Fr\"{o}jdh and H. Johannesson, Phys. Rev. Lett. {\bf 75}, 300(1995).
\bibitem{12}Y. Wang, J. Voit and F.-C. Pu, Phys. Rev. B {\bf 54}, 8491(1996).
\bibitem{13}N. Andrei and H. Johannesson, Phys. Lett. {\bf A100}, 108(1984).
\bibitem{14}G. Bed\"{u}rftig, F.H.L. E{\ss}ler and H. Frahm, Phys. Rev. Lett. {\bf 77}, 5098(1996).
\bibitem{15}Y. Wang, J. Dai, Z. Hu and F.-C. Pu, Phys. Rev. Lett. {\bf 79}, 1901(1997).
\bibitem{16}Y. Wang and J. Voit, Phys. Rev. Lett. {\bf 77}, 4934(1996);(E) {\bf 78}, 3799(1997).
\bibitem{17}Y. Wang, Phys. Rev. B {\bf 56}, 14045(1997).
\bibitem{18}E.K. Sklyanin, J. Phys. A {\bf 21}, 2375(1988); C. Destri and H.J. de Vega, Nucl. Phys. B{\bf 361}, 361(1992);
B {\bf 374}, 692(1992).
\bibitem{19}P. Schlottmann, Phys. Rev. B {\bf 36}, 5177(1987).
\bibitem{20}A. Foester and M. Karowski, Nucl. Phys. B {\bf 396}, 611(1993); B {\bf 408}, 512(1993).
\bibitem{21}C.N. Yang, Phys. Rev. Lett. {\bf 19}, 1312(1967).
\bibitem{22}Note here the scattering matrix is the inverse of that defined in ref.15.
\bibitem{23}Due to the reflection symmetry, $q$ and $-q$ correspond to the same state. It can be shown that when two modes have
the same absolute value, $|q_j|=|q_l|$ or $|\lambda_\alpha|=|\lambda_\beta|$, the wave function is zero. That means such situations
are forbidden. This is a common feature of the 1D open boundary systems.
\bibitem{24}C.J. Hamer, G.R.W. Quisel and M.T. Batchelor, J. Phys. A {\bf 20}, 5677(1987); M.T. Batchelor
and C.J. Hamer, J. Phys. A {\bf 23}, 761(1990); F.C. Alcaraz, M.N. Barber and M.T. Batchelor, Ann. Phys.
(N.Y.) {\bf 182}, 280(1988); A. Asakawa and M. Suzuki, J. Phys. A {\bf 29}, 225(1995).
\bibitem{25}L.D. Faddeev and L.A. Takhtajan, Phys. Lett. {\bf A85}, 375(1981).
\bibitem{26}N. Andrei and C. Destri, Phys. Rev. Lett. {\bf 52}, 364(1984).
\bibitem{27}Y. Wang and U. Eckern, cond-mat/9805332.
\bibitem{28}N. Andrei, K. Furuya and J. Lowenstein, Rev. Mod. Phys. {\bf 53}, 331(1983).
\bibitem{29}C.N. Yang and C.P. Yang, J. Math. Phys. {\bf 10}, 1115(1969).
\bibitem{30}A.G. Izergin, V.E. Korepin and N. Yu. Reshetikhin, J. Phys. A {\bf 22}, 2615(1989).
\bibitem{31}J. Carmelo and A.A. Ovchinnikov, J. Phys. C {\bf 3}, 757(1991).
\bibitem{32}M. Takahashi, Prog. Theor. Phys. {\bf 46}, 1388(1971).
\bibitem{33}P. Nozi\`{e}res, J. Low. Temp. Phys. {\bf 17}, 31(1974).


\end{references}
\end{document}